\documentclass[mypaper,7pt,twoside]{CoAst}
\usepackage{epsf,graphicx,fancyhdr}
\usepackage{url}
\renewcommand{\chaptermark}[1]{\markboth{#1}{}}

\renewcommand{\headrulewidth}{0pt}
\newcommand{\kopf}{\small\itshape Comm. in Asteroseismology \\ Vol. number, publication date (will be inserted in the production process)}

\newcommand{\Authors}[1]{\begin{center}\normalsize\bf\sf #1 \end{center}}

\renewcommand{\author}[1]{\begin{center}\normalsize\bf\sf #1 \end{center}}
\newcommand{\Address}[1]{\begin{center}\small\sf #1 \end{center}}

\renewenvironment{abstract}{\section*{Abstract}\normalsize\sf}{}
\newcommand{\References}[1]{\begin{flushleft}{\large References\\}\vspace*{2mm}\small #1 \end{flushleft}}

\newcommand{\chapterCoAst}[2]{\chapter[\sf\normalsize #1\\ \footnotesize \hspace*{5mm}by #2 \sf\normalsize][]{#1\\}\rhead[\fancyplain{}{\sf\footnotesize \center{#1}}]{\fancyplain{}{\sffamily\thepage}}\lhead[\fancyplain{\kopf}{\sffamily\thepage}]{\fancyplain{\kopf}{\sf\footnotesize \center{#2}}}}

\renewcommand{\chaptername}{}

\def\rfr{\smallskip\par\noindent
        \hangindent=7truemm
        \hangafter=1}

\begin{document}
\sf

\chapterCoAst{HELAS Local Helioseismology Activities}
{H.\,Schunker \& L.\,Gizon} 
\Authors{H.\,Schunker, L.\,Gizon} 
\Address{Max-Planck-Institut f\"ur Sonnensystemforschung, Max-Planck-Strasse 2, 37191, Katlenburg-Lindau, Germany\\
}

\noindent
\begin{abstract}
The main goals of the HELAS local helioseismology network activity are to consolidate this field of research in Europe, to organise scientific workshops, and to facilitate the distribution of observations and data analysis software. Most of this is currently accomplished via a dedicated website  at  \url{http://www.mps.mpg.de/projects/seismo/NA4/}. In this paper we list the outreach material, observational data, analysis tools and modelling tools currently available from the website and describe the focus of the scientific workshops and their proceedings.

\end{abstract}


\section{Introduction}
Local helioseismology studies the three dimensional structure of sunspots and active regions, local mass flows and enables us to identify magnetic activity on the far-side of the Sun. Many properties of the solar interior that may be probed with local helioseismology have yet to be revealed. At the moment there exists  abundant high quality helioseismic data from the Global Oscillation Network Group (GONG) and the Michelson Doppler Imager (MDI) instrument to work with. In the near future the Solar Dynamics Observatory (SDO) will be launched with the Helioseismic and Magnetic Imager (HMI) on board, which will collect full-disk, high-resolution data. To support European local helioseismology and to exploit the large amounts of data efficiently it is helpful to select and make available useful observations and analysis tools.

The European Helio- and Asteroseismology Network (HELAS)  was created to co-ordinate the exchange of knowledge, data and software tools amongst  researchers. There are four main network activities. One of these network activities (NA4) carries out the objectives of HELAS specifically for local helioseismology. It is implemented via a dedicated website and through meetings and workshops.

In the following section we discuss the website which holds about 1 TB of data and provides a platform for the exchange of observations, analysis tools, solar models and general information  about local helioseismology. The  workshops and meetings are described towards the end of this paper. Finally, we list the publications pertaining specifically to HELAS local helioseismology.

\section{The HELAS local helioseismology website}\label{www}

The address of the HELAS local helioseismology website is \url{http://www.mps.mpg.de/projects/seismo/NA4/} and is hosted by the Max Planck Institute for Solar System Research (MPS), in Katlenburg-Lindau, Germany. It boasts straightforward access to specially selected useful and interesting helioseismic data sets, outreach material,  tools to analyse the observations and modelling codes suitable for local helioseismology. Since September 2007 there have been 380 absolute unique visitors from 64 countries.  The website is regularly updated with new material when it becomes available. The system is backed up regularly, ensuring ongoing availability of the data from the website. This section goes through the main parts of the website.

\begin{figure}
\includegraphics[width=0.99\textwidth]{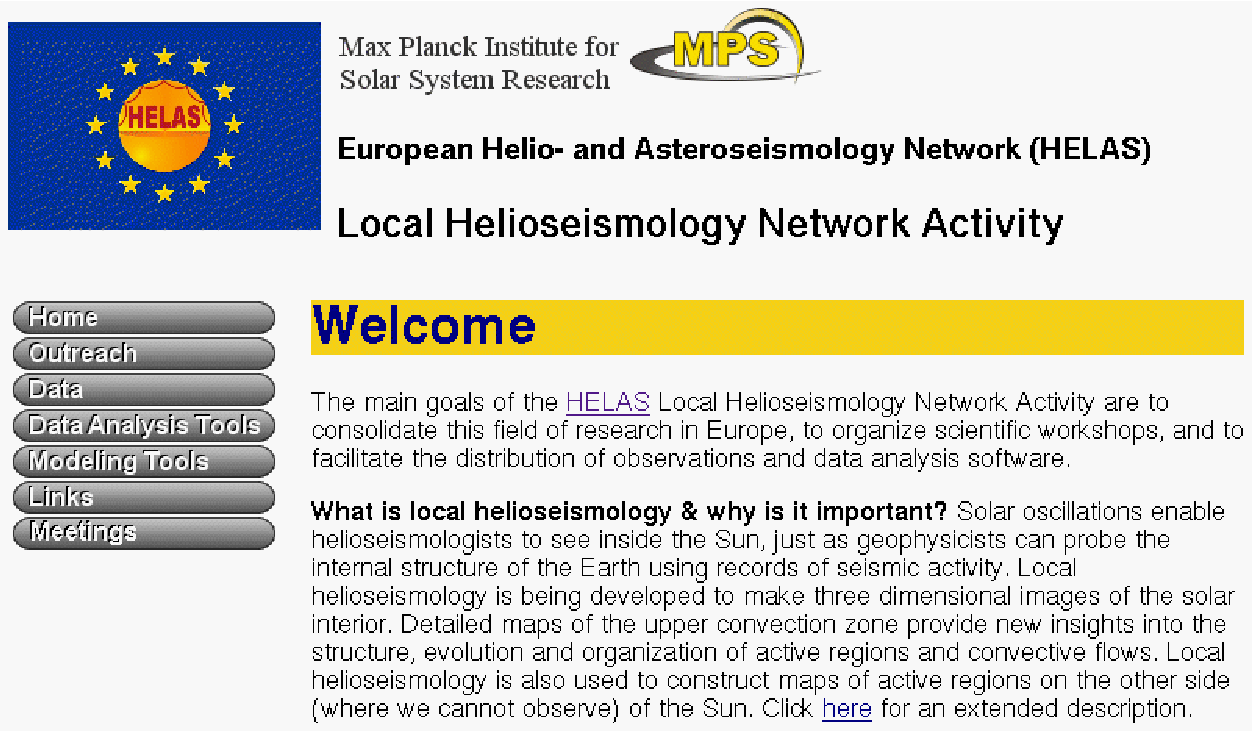}
\caption{Homepage of the HELAS local helioseismology website which acts as a platform for the exchange of knowledge, data, analysis tools and modelling tools at  \texttt{http://www.mps.mpg.de/projects/seismo/NA4/}.}
\label{homepage}
\end{figure}

\subsection{Documentation and outreach material}
The website is aimed at both students and professional scientists in the field. We provide links to educational material comprising media coverage, technical reports and scientific papers at \url{http://www.mps.mpg.de/projects/seismo/NA4/helasNA4_General.html}. The media coverage includes science magazines and news reports in a variety of languages. The technical information incorporates review papers, a list of appropriate text books, lecture notes and  PhD theses. We also link to the homepages of academic groups active in the field. The outreach section should provide a scientist or a student with enough information to understand the basics of helioseismology and itÕs importance.

\subsection{Observations}\label{obs}

\begin{figure}[tbh]
\begin{center} 
\includegraphics[width=0.99\textwidth]{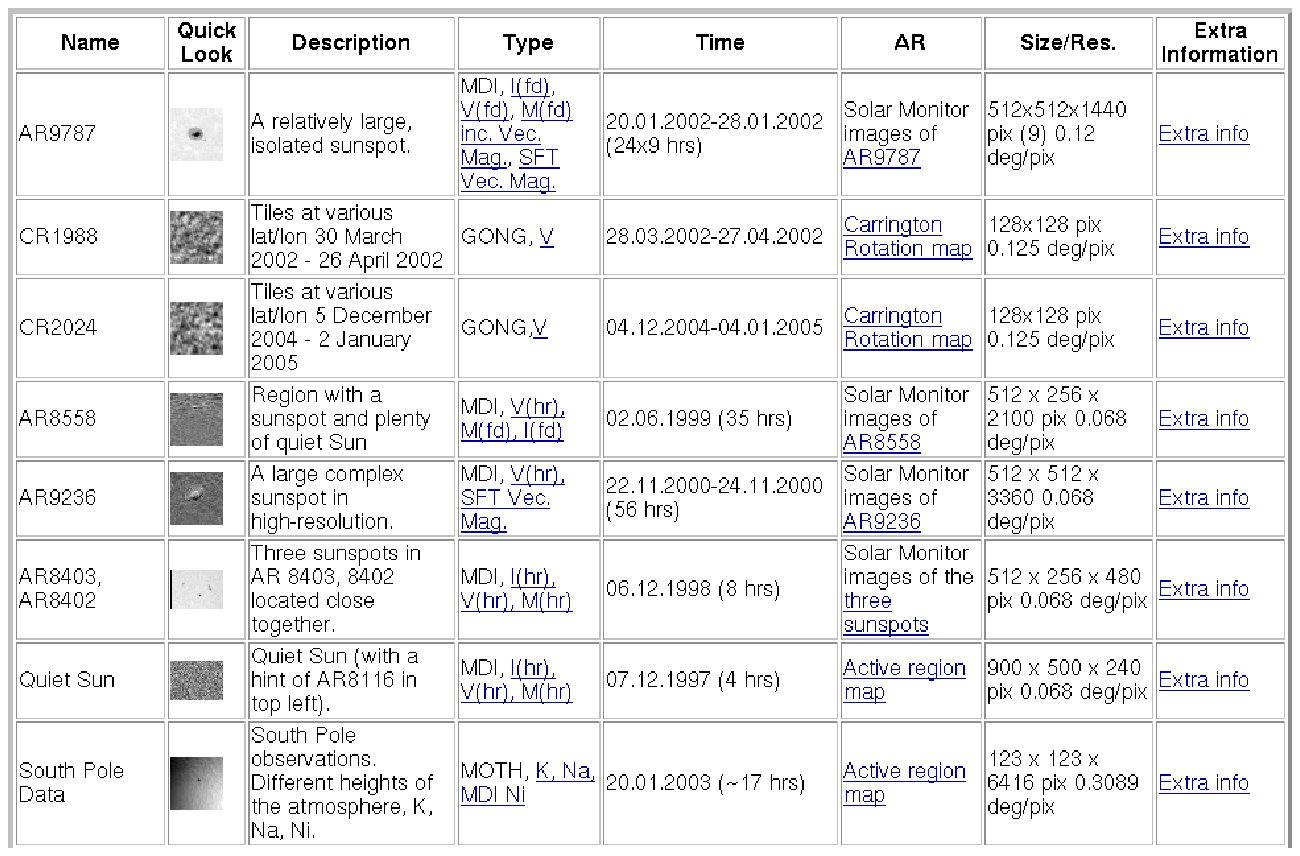}
\caption{Quick look data table of some of the available data sets  \texttt{http://www.mps.mpg.de/projects/seismo/NA4/DATA/data\_access.html}.}
\label{datatable}
\end{center}
\end{figure}

One of the purposes of HELAS local helioseismology is to select useful observational data sets and make them available. In this section the selected data sets that are currently available at \url{http://www.mps.mpg.de/projects/seismo/NA4/DATA/data_access.html} are listed. An earlier description can be found in Schunker~et\,al.~(2008). The observations provided on the website are predominantly  Dopplergrams, magnetograms, intensity continuum images and vector magnetograms from the Solar and Heliospheric Observeratory's (SOHO)  MDI, the GONG and the Magneto-Optical filter at Two Heights (MOTH) instruments in Flexible Image Transport System (FITS) format. The observations cover entire Carrington rotations as well as selected active and quiet Sun regions. The observational data sets are listed in a quick look format including a sample graphic and a description of the data  as shown in the data table in Figure~\ref{datatable}. We provide a dedicated web page for each data set with more detailed information relevant to the particular observations. 

The most complete data set available from the HELAS local helioseismology website are the observations of AR9787.  The data consists of nine full days of MDI Doppler velocity and magnetic field observations of the large, isolated sunspot in AR9787 with a cadence of one minute. One intensity continuum image exists for each six hours. Solar Flare Telescope (SFT) vector magnetograms are also available in this data set. This data set was analysed extensively at the second HELAS local helioseismology workshop in January 2008 and the results are published in Gizon~et\,al.~(2008). 

The rest of the data sets available from the HELAS local helioseismology website are now listed: (1) Doppler velocity observations from the GONG mapped ÔtilesÕ throughout Carrington Rotations 1988  and 2024;  (2) MDI high resolution, thirty-five hour Doppler velocity observations of AR8555 and a portion of quiet Sun as well as full-disk intensity and magnetograms; (3) Observations of a large, complex sunspot, AR9236, observed with MDI high-resolution for 56 hours is accompanied by SFT vector magnetograms; (4) MDI high-resolution Doppler, magnetic and intensity continuum observations of three sunspots in close proximity observed over eight hours; (5) Four hours of MDI high-resolution Doppler velocity observations of quiet Sun; (6) Multiple spectral line observations from the MOTH instrument provide unique simultaneous multi-height observations of the same spatial region; (7) AR9026 MDI data covers the time of an X2.3 class solar flare which has been reported to cause a sunquake; (8) A series of active regions and close quiet Sun regions analysed identically are available for comparative purposes. 

\begin{figure}[htb]
\begin{center} 
\includegraphics[width=1\textwidth]{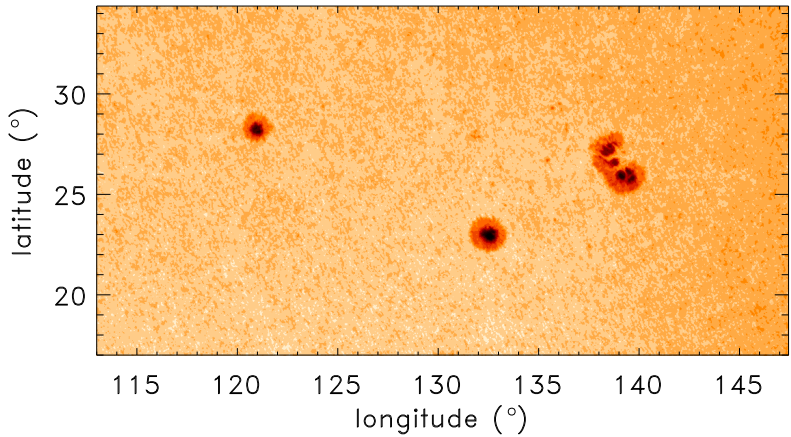}
\caption{MDI continuum image from a particular HELAS data set showing three sunspots in active regions AR8402 and AR8403. A total of 8 hours of high-resolution MDI Dopplergrams are available for helioseismic analysis.}
\label{dataex}
\end{center}
\end{figure}

 Figure~\ref{dataex} shows a quick look intensity image of an example data set with three sunspots in AR8402 and AR8403. A total of 8 hours of high-resolution MDI Dopplergrams are available for local helioseismology. For each set of active region observations a link is provided to the Solar Monitor webpage with quick look maps in various observing spectral lines. External links to active region and synoptic maps provide a good overview of the solar activity at the time.

Important information has been collated for each data set:  the Carrington latitude and longitude of the observation, links to sample headers, FITS header keyword definitions, and general information on the physical active region characteristics. These characteristics include the co-ordinates of the sunspot, the magnetic type, classification, the area, the longitudinal extent of the sunspot, the average umbral and penumbral boundaries and the number of sunspots present on the solar disk at the time of the observation.  In some cases far-side images of the active region are shown which give an indication of the active region's life span.  This is important for a full analysis of the observations.

\subsection{Data reduction}

Most of the observations have been pre-processed and it is important to know exactly how. The webpage \url{http://www.mps.mpg.de/projects/seismo/NA4/DATA/data_red.html} links to published papers about the instruments, the data processing pipelines and the known characteristics of the data. Papers describing the merging of GONG data and the onsite data reduction are available for direct download. A schematic of the pipeline involved in delivering GONG data products is linked. The paper by Scherrer~et\,al.~(1995) contains all the relevant instrument details for the MDI observations. The MDI website contains many more details in regards to image quality,  duty cycle and other performance characteristics.  Similar information for the other instruments is  primarily available in the form of papers.

Most of the observations available have been further processed. In local helioseismology, tracking of the local solar surface in time is necessary to remove solar rotation. A particular latitudinal rotation profile is used to do the tracking and we provide each profile for each data set. In the majority of cases for the data provided by the website this is done using an azimuthal-equidistant projection and mathematical descriptions of the many possible projections that could be used are linked.

\subsection{Data analysis tools}

Currently, three data analysis software tools are available from the HELAS local helioseismology activity website at \url{http://www.mps.mpg.de/projects/seismo/NA4/SW/}: the ring diagram analysis pipeline,  code to measure travel times and holography codes. Documentation for the codes is provided in the form of linked reports, papers and presentations. The actual computer codes are available for direct download. Here we describe the data analysis software tools available.

\subsubsection{Data analysis tool 1: Ring diagram pipeline}
Ring diagram helioseismology (Hill~1988) is based on the analysis of the local acoustic power spectra in wavenumber and frequency space. In planes of constant frequency the power is concentrated in rings. Analysing the changes in the shape of these rings provides information about anomalies in the sub-surface structure of the Sun, such as flows and sound speed perturbations.

The GONG ring diagram software package comprises all the binaries, scripts and support files needed to obtain the subsurface averaged velocity flows associated with individual sections over the solar surface using the ring diagram technique. It has been optimised to take GONG Dopplergrams as input, although it could easily be adapted to accept other helioseismic observations. Various authors wrote the individual parts of the pipeline, however the package was put together for HELAS by I. Gonz\'alez-Hern\'andez (National Solar Observatory) and has been made available to download from the HELAS local helioseismology web page. Extensive documentation related to installing and running the pipeline is also linked, courtesy of A. Zaatri (Kiepenheuer-Institut f\"ur Sonnenphysik) and T. Corbard (Observatoire de la C\^ote d'Azur), as well as presentations and relevant scientific papers.

\subsubsection{Data analysis tool 2: Helioseismic holography codes}
Helioseismic holography (Lindsey \& Braun~2000) takes the acoustic amplitudes observed at the surface and computationally regresses them into the interior either forward in time (ingression) or backward in time (egression). The results give us an acoustic image of the solar interior.

The basic codes were written by C. Lindsey (Colorado Research Associates) to calculate the ingression and egression  of pre-processed MDI Dopplergrams for helioseismic holography and are available to download. The HELAS holography webpage  gives the specific information required by these programs so that users can tailor their input data accordingly. The web site also provides documentation about the technique by providing links to the scientific papers on the basic principles of helioseismic holography.

\subsubsection{Data analysis tool 3: Fitting travel times}
The goal of time-distance helioseismology is to calculate the travel time of wave packets observed on the surface over some distance. This is done by calculating the cross-covariance function between the signal at two points. The travel times are measured by fitting wavelets to the cross-covariance.

The codes that are made available on the HELAS local helioseismology website consist of IDL routines for measuring travel times from the cross covariance. Two methods are implemented: a Gabor-wavelet fit and a one-parameter fit.  Example cross-covariances and instructions for running the code are included in the package. A full description of the fitting methods is given in Roth~et\,al. (2007).

\subsection{Modelling tools}\label{hilev}
Models and simulations are important for understanding and interpreting the results of helioseismic analysis. We have collated a selection of model and simulation source codes relevant for local helioseismology. Here we describe the modelling tools that are available from \url{http://www.mps.mpg.de/projects/seismo/NA4/MODEL/}.

\subsubsection{Modelling tools 1: Numerical simulation of wave propagation}

The Semi-spectral Linear MHD (SLiM) code numerically simulates the propagation of linear waves through an arbitrary three-dimensional atmosphere (Cameron ~et\,al. ~2007). From these simulations the interaction of acoustic waves with various inhomogeneities in the solar atmosphere can be studied.

\begin{figure}[tbh]
\begin{center} 
\includegraphics[width=0.99\textwidth]{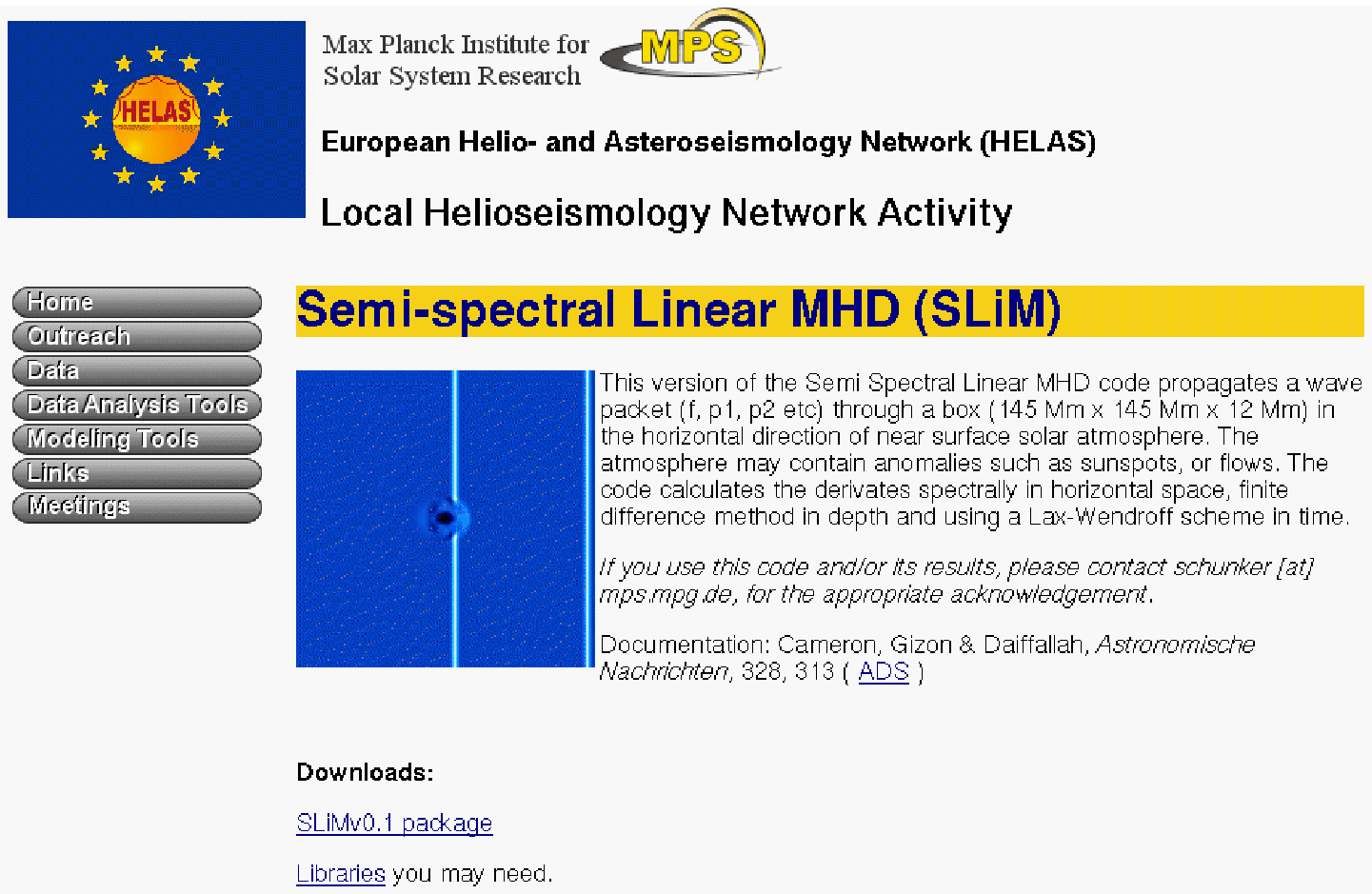}
\caption{Screen grab of  the ÒSLiMÓ higher level application tool webpage at \texttt{http://www.mps.mpg.de/projects/seismo/NA4/MODEL/SLiM.html}. }
\label{slim}
\end{center}
\end{figure}

The code is written in Fortran90 and documented by published papers. Instructions to run the code comes in a READ\_ME file in the download package. Fortran libraries required to run the code are also provided. See the screen shot in Figure~\ref{slim} for more details.

\subsubsection{Modelling tools 2: Sunspot \& fluxtube models}

The interaction of acoustic waves with magnetic inhomogeneities of the solar atmosphere is an important subject of research. Codes which model various magnetic structures, sunspots and flux tubes, are made available from the website.

An IDL code creates a  typical sunspot structure in a solar like atmosphere (Khomenko \& Collados~2006) . The variable parameters are the magnetic field strength, the radius of the sunspot and the Wilson depression. A READ\_ME file describes how to run the code and further information can be obtained by reading the listed publications.

A Fortran code that calculates magnetohydrostatic flux tubes according to the method of Steiner~et\,al.~(1986) is also provided. One input file contains all of the variable parameters which is easily modified. Extensive documentation to run the code is provided in the READ\_ME file.

\subsubsection{Modelling tools 3: Born travel-time sensitivity kernels}

Travel-time sensitivity kernels are necessary to interpret helioseismic inversions. They describe the region of the Sun that a particular wave packet is sensitive to in the presence of some perturbation. In this case, the code to calculate the kernels is not supplied directly as there are numerous input parameters. Instead, Y. Saidi has developed a web interface and pipeline to expedite the calculation process. The computing resources are  located at the MPS. 

\begin{figure}
\includegraphics[width=0.99\textwidth]{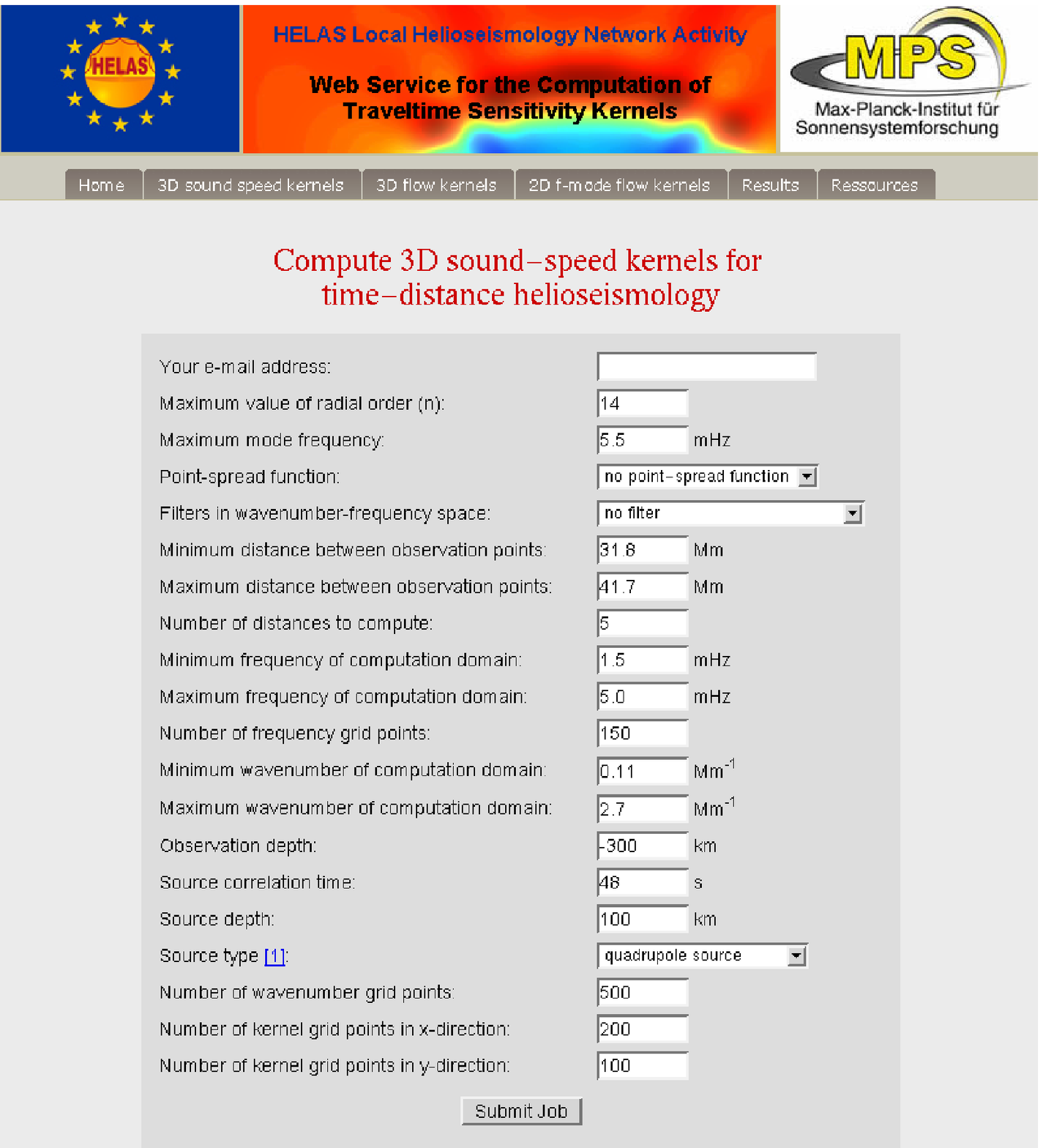}
\caption{Screen grab of the web interface  showing some of the input parameters to calculate 3D sound speed kernels for the computation of travel-time sensitivity kernels. The interface is linked from \texttt{http://www.mps.mpg.de/ projects/seismo/NA4/MODEL/tt\_interface.html}.}
\label{wi}
\end{figure}

The interface allows users to specify input parameters for the kernel calculation (see Figure~\ref{wi}). The linear sensitivity kernels are based on a single-scattering Born approximation. Users have the choice between three different types of calculations as seen in Figure~\ref{wi}. The first set of calculations provides three- dimensional p-mode kernels for sound-speed perturbations (Birch~et\,al.~2004). The second set of calculations provides three- dimensional travel-time sensitivity kernels for flows (Birch \& Gizon~2007). The third set provides two-dimensional f-mode kernels for flows (Jackiewicz~et\,al.~2007). Additional references are listed on the website for each calculation.

The results of the submitted calculation are available to download from a web address that is e-mailed to the user. An email is sent at each step of the computation. The first one will inform the user that the request has been received, the second one will notify the user that the computation has started and the third email will notify the user that the computation has ended, and if it is successful it will also carry the web address to download the results of the computation.  Yacine Saidi developed this web interface and relevant details can be found in his Master's thesis (Saidi 2006).

\subsubsection{Modelling tools 4:  Ray tracing code}

Time-distance helioseismology measures and interprets the travel times of wave packets propagating between two points located on the solar surface. The travel times are then inverted to infer sub-surface properties that are encoded in the measurements. Thus, it is useful to have some knowledge of the path the wave packet takes and the region of the Sun it passes through.

Codes are provided that calculate helioseismic ray paths written by A.C. Birch (Colorado Research Associates). There are three main codes which calculate the group time as a function of distance, the ray paths, the phase and group times along the path (Birch~2002). Documentation explaining the execution and function of the codes is provided.

\section{Meetings and workshops}\label{meet}

There have been two HELAS workshops and one splinter meeting dedicated to the HELAS local helioseismology network activity, all of which had a high international participation. The first workshop, \textit{Roadmap for European Local Helioseismology} (\url{http://www.oca.eu/HELAS/}), was held at the Observatoire de la C\^ote d'Azur, Nice, France on 25-27 September 2006.  There were 38 participants  who gave talks on the status of local helioseismology. In light of this there were discussions about the direction local helioseismology should take, particularly in Europe. The proceedings consist of 28 papers and were published in the journal \textit{Astronomische Nachrichten}  Volume 328, Issue 3-4 accessible from the web site \url{http://www3.interscience.wiley.com/journal/114173529/issue}.

The second workshop  was held at the Kiepenheuer-Institut f\"ur Sonnenphysik in Freiburg, Germany on 7-11 January  2008 (\url{http://www.mps.mpg.de/projects/seismo/HLHW2/}).  Twenty-eight people attended by invitation and were divided into three teams, that covered the analysis of active region AR9787, numerical MHD simulations, and preparations for the Solar Dynamics Observatory. During the workshop Team 1 actively analysed the sunspot in AR9787 using MDI/SOHO observations made available on the HELAS local helioseismology website. Three main methods of helioseismology -- time-distance, ring-diagram and helioseismic holography -- were involved. This made it possible to directly compare the results for identical observations by eliminating any possible differences in initial data reduction. This had not been done before for a sunspot.  
Team 2 discussed  issues related to the numerical treatment of wave propagation in the near surface layers of the Sun. Particular emphasis was placed on two topics: the need for a set of standard tests which numerical codes should reproduce as a form of code validation; and on the various ways to construct background solar models containing inhomogeneities.
Team 3 made significant progress towards handling the large amounts of data from the Solar Dynamics Observatory (SDO) which will be launched in 2009. The team focused on implementing and discussing the Data Record Management System (DRMS), which is the software that will be responsible for managing SDO data. The DRMS is being developed and distributed by the Joint Science Operations Centre (JSOC) at Stanford University in the USA, and is also operational at the German Data Centre (GDC) for SDO. 
The analysis of AR9787 lead to interesting discussions which were continued at the International Space Science Institute meeting, \textit{The origin and dynamics of solar magnetism} that was held in Bern, Switzerland on 21-25  January  2008. This has resulted in a forthcoming publication in Space Science Reviews (Gizon~et\,al.~2008).

A splinter session devoted to discussing the HELAS local helioseismology deliverables was a part of the HELAS II International Conference - \textit{Helioseismology, Asteroseismology and MHD Connections} (\url{http://www.mps.mpg.de/meetings/seismo/helas2/}) held in G\"ottingen, Germany on 20-24 August  2007.  The suggestions and comments from the participants contributed to the selection of observations, software tools and models made available on the HELAS local helioseismology website. The proceedings of this meeting are published in the \textit{Journal of Physics: Conference Series} volume 118.
In collaboration with the SOHO 19/ GONG 2007 meeting held in Melbourne, Australia  on 9-13 July 2007, a Topical Issue of \textit{Solar Physics} was also published and is listed in the following section. This Topical Issue places a strong emphasis on local helioseismology.

Judging from the resulting publications, the workshops fostered strong collaborations between the working groups active in local helioseismology in Europe and elsewhere around the world. 

\section{Publications}
 A total of 163 papers have resulted from the HELAS local helioseismology network activity. These papers appear in the following publications:
\begin{itemize}
\item \textit{Astronomische Nachrichten}, Volume 328, Issue 3-4. Proceedings from the First HELAS Workshop, \textit{Roadmap for Local Helioseismology} held at the Observatoire de la C\^ote d'Azur, Nice, France on 25-27 September  2006. The volume consists of 28 papers.\\

\item \textit{Solar Physics}, `Helioseismology, Asteroseismology and MHD Connections', volume 251, 2008. 
This is a joint effort between the HELAS II International Conference   and  the SOHO 19/ GONG 2007 meeting. Many of the papers included here represent work presented at one or other of the meetings,
but this Topical Issue was opened for general submission on their core topics.  From the 43 papers in the volume, 26 are focused on local helioseismology.

\item \textit{Journal of Physics: Conference Series, Proceedings of the second HELAS international conference: Helioseismology, Asteroseismology and MHD Connections}, volume 118, 2008.  The volume contains 11 local helioseismology papers out of 91 papers in total.
\end{itemize}

An additional paper has been accepted for publication:
\begin{itemize}
\item L. Gizon, H. Schunker, C.S. Baldner, S. Basu, A.C. Birch, R.S. Bogart, D.C. Braun, R. Cameron, T.L. Duvall, Jr, S.M. Hanasoge, J. Jackiewicz, M. Roth, T. Stahn, M.J. Thompson, S. Zharkov, `Helioseismology of sunspots: A case study of NOAA region 9787', \textit{Space Science Reviews}, submitted, 2008. This is a joint publication resulting from the second HELAS local helioseismology workshop.
\end{itemize}

\section{Future activity}

One more workshop  is planned to be held in the first half of 2009. The observations, software analysis tools and modelling code will continue to be modified and updated to keep up with progressions in the field.

\section{Acknowledgements}
Many thanks to all contributors to the HELAS local helioseismology network activity including Charles Baldner, Aaron Birch, Rick Bogart, John Bolding, Doug Braun, Robert Cameron, Thierry Corbard, Tom Duvall, Wolfgang Finsterle, Shravan Hanasoge, Irene Gonz\'alez-Hern\'andez, Frank Hill, Jason Jackiewicz, Elena Khomenko, A.G. Kosovichev,  John Leibacher, Charles Lindsey, Markus Roth, Yacine Saidi, Oskar Steiner, Mike Thompson, Thomas Wiegelmann, Amel Zaatri and Sergei Zharkov.
Special thanks to all others  from the HELAS local helioseismology nodes at the Observatoire de la C\^ote d'Azur, Max-Planck-Institut f\"ur Sonnensystemforschung, Kiepenheuer-Institut f\"ur Sonnenphysik and the University of Sheffield for their support.


\References{
\rfr Birch, A.C.  2002, PhD thesis
\rfr Birch, A. C., Kosovichev, A. G.\& Duvall, T. L. 2004, Astrophys. J. 608, 580
\rfr Birch, A. C. \& Gizon, L. 2007, Astron. Nachr. 328, 228
\rfr Cameron, R., Gizon, L., \& Daiffallah, K. 2007 Astron. Nachr., 328, 313
\rfr Duvall, T.L., Jefferies, S.M., Harvey, J.W.  \& Pomerantz, M.A.  1993, Nature, 362, 430
\rfr Gizon, L., Schunker, H.,  Baldner, C.S.,  Basu, S., Birch, A.C.,  Bogart, R.S., Braun, D.C., Cameron, R., Duvall, Jr, T.L., Hanasoge, S.M., Jackiewicz, J., Roth, M., Stahn, T.,  Thompson, M.J., Zharkov, S. 2008, Space Sci. Rev., accepted 
\rfr Hill, F. 1988, Astrophys. J., 333, 996
\rfr Jackiewicz, J., Gizon, L., Birch A. C., \& Duvall, T. L. 2007, Astrophys. J. 671, 1051
\rfr Khomenko, E. \& Collados, M. 2006, Astrophys. J., 653, 739
\rfr Lindsey, C. \& Braun, D.C. 2000, Sol. Phys., 192, 261
\rfr Roth, M., Gizon, L., \& Beck, J.G.  2007, Astron. Nachr., 328, 215
\rfr Saidi, M.Y. 2006, M.Sc. Dissertation, Universit\'es Paris Sud XI, Paris
\rfr Scherrer, P. H., Bogart, R. S., Bush, R. I., Hoeksema, J. T., Kosovichev, A. G., Schou, J., Rosenberg, W., Springer, L., Tarbell, T. D., Title, A., Wolfson, C. J., Zayer, I. \& MDI Engineering Team 1995, Sol. Phys., 162, 129
\rfr Schunker, H., Gizon, L. \& Roth, M. 2008 Proceedings of the Second HELAS International Conference, J. Phys. Conf. Ser., vol. 118, doi:10.1088/1742-6596/118/1/012087
\rfr Steiner, O., Pneuman, G.W. \& Stenflo, J.O. 1986 Astron. Astrophys. 170, 126

}

\end{document}